\documentclass[aps,prd,twocolumn,floatfix,nofootinbib,superscriptaddress]{revtex4}
\usepackage{dcolumn}

\usepackage{amsmath,amsfonts,extarrows,amssymb,mathrsfs,times,bm}
\usepackage{extarrows}
\usepackage{graphicx}
\usepackage{float}
\usepackage{graphicx,epstopdf}
\usepackage{dcolumn}
\usepackage{exscale}
\usepackage{relsize}
\usepackage{mathtools}
\usepackage[colorlinks=true,breaklinks=true,linkcolor=blue,citecolor=blue,urlcolor=blue]{hyperref}
\usepackage{tikz}
\usepackage[usenames,dvipsnames]{pstricks}
\usepackage{subfigure}
\usepackage{epsfig}

\newcommand{\nn}{\nonumber}

\begin{document}

\title{Deflection of charged massive particles by a four-dimensional charged Einstein-Gauss-Bonnet black hole}
\author{Zonghai Li}
\affiliation{Center for Astrophysics, School of Physics and Technology, Wuhan University, Wuhan 430072, China}
\author{Yujie Duan}
\affiliation{Center for Astrophysics, School of Physics and Technology, Wuhan University, Wuhan 430072, China}
\author{Junji Jia}
\email[Corresponding author: ]{ junjijia@whu.edu.cn}
\affiliation{MOE Key Laboratory of Artificial Micro- and Nano-structures \& Center for Astrophysics, School of Physics and Technology, Wuhan University, Wuhan, 430072, China}
\date{\today}

\begin{abstract}
Based on the Jacobi metric method, this paper studies the deflection of a charged massive particle by a novel four-dimensional charged Einstein-Gauss-Bonnet
black hole. We focus on the weak field approximation and consider the deflection angle with finite distance effects. To this end, we use a geometric and topological method, which is to apply the Gauss-Bonnet theorem to the Jacobi space to calculate the deflection angle. We find that the deflection angle contains a pure gravitational contribution $\delta_g$, a pure electrostatic $\delta_c$ and a gravitational-electrostatic coupling term $\delta_{gc}$. We also show that the electrostatic contribution $\delta_c$ can also be computed by the Jacobi metric method using the GB theorem to a charge in a Minkowski flat spacetime background. We find that the deflection angle increases(decreases) if the Gauss-Bonnet coupling constant $\alpha$ is negative(positive). Furthermore, the effects of the BH charge, the particle charge-to-mass ratio and the particle velocity on the deflection angle are analyzed. 
\end{abstract}

\maketitle

\section{Introduction}

The gravitational deflection of light is one of the classical tests of General Relativity (GR). It was first verified in 1919~\cite{DED1920,Will2015}, and now has developed into a very active area of research, namely, the gravitational lensing (GL). The analysis of the signatures of the strong/weak GL will help us understanding the properties of the source, the lens and the messenger. Due to the importance of this fact, GL has become an important 
window for exploring the unknown world and has been widely used in cosmology and astronomy~\cite{Hoekstra2013,Brouwer2018,Bellagamba2019,Vanderveld2012,cao2012,zhanghe2017,Huterer2018,SC2019,Andrade2019}. In addition to light, on the other hand, we are also interested in deflection and GL of other signals such as gravitational wave and timelike particle like neutrinos. Apart from the mass, 
another fundamental property of a particle is its charge. For massive neutral particles (NMP), their GL has attracted great attention~\cite{AR2002,AP2004,Bhadra2007,Yu2014,He&lin2016,He2020,Liu&Jia2016,Pang&Jia2019,Jiaepjc2020}. However, only a few researchers considered the deflection effect of charged massive particle (CMP) in previous works~\cite{CGV:charged,Jusufi:cmp}. One of the reasons may be that the motion of CMP is more complicated than that of NMP, because the path of CMP is no longer a geodesic in a charged spacetime. However, the deflection of CMP is meaningful and worthy of study from both the theoretical and observational points of view. 

Although GR is very beautiful, it is not the final theory of gravity. The modern view is that GR is just a low-energy effective field theory and we need to go beyond it. The Einstein-Gauss-Bonnet (EGB) theory is one of the most promising candidates for modified theories of gravity. It can be derived from the low energy limit of string theory~\cite{Zwiebach,Garraffo}. In EGB theory, when spacetime dimension $D=4$, the Gauss-Bonnet term is a total derivative, so it does not contribute to the gravitational dynamics. However, Glavan and Lin recently proposed a non-trivial four-dimensional EGB theory of gravity~\cite{GlavavLin}. They rescaled the coupling constant $\alpha\to  \alpha/(D-4)$ and the four-dimensional theory is defined as a limit of $D\to  4$. Interestingly, this novel four-dimensional EGB theory bypasses the Lovelock's theorem and avoids Ostrogradsky instability. In addition, a static spherically symmetric (SSS) black hole (BH) solution in this theory was obtained in Ref.~\cite{GlavavLin}. Since then, a number of exact solutions have been 
found in this four-dimensional EGB theory, such as charged AdS BH~\cite{Fernandes2020}, rotating BH~\cite{WeiLiu,KumarGhosh}, Bardeen BH~\cite{KumarKumar}, Hayward BH~\cite{KumarGhoshHayward}, Born-Infeld BH~\cite{YangKe}, nonlinear magnetically charged BH~\cite{Jusufi:nmcbh}, Morris-Thorne wormholes~\cite{Jusufi:mewha}, traversable thin-shell wormhole~\cite{Liu:ttsw}, and so on. Moreover, the approach of Glavan and Lin in Ref. \cite{GlavavLin} was generalized to Einstein-Lovelock gravity~\cite{KonoplyaEL,KonoplyaZhidenko,CasalinoColleaux}. 
Meanwhile, the rationality of the definition of the four-dimensional EGB theory was discussed, which greatly increased the research efforts of the theory~\cite{GSismanTekin,HennigarPollack,FernandesCarrilho,Bonifacio,Mahapatra,Aoki2020}. The quasinormal modes~\cite{Konoplya2003,MSChurilova1,MSChurilova2,AkashMishra,Zhangsanfeng}, thermodynamics~\cite{SAHMansoori,ShuxuanYing,DVSingh1,DVSingh2,YYWang}, BH shadows~\cite{GuoLi2020,ZengHQ} were investigated in the context of the four-dimensional EGB theory. Naturally, the GL in this novel four-dimensional EGB gravity theory was also studied by some authors. Islam et al.  investigated the GL of SSS BH in the strong and weak deflection limits~\cite{Islam}. Kumar et al.  studied the deflection of light by a charged BH~\cite{Kumar2020}. Jin et al.  studied the strong GL of SSS BH surrounded by unmagnetised plasma medium~\cite{Jin}. Heydari-Fard et al.  calculated the bending angle of light in dS spacetime using Rindler-Ishak method~\cite{Heydari-Fard}. Panah et al.  computed the deflection angle of light from the charged AdS BH~\cite{PanahkJafarzade}. Jafarzade et al.  considered the lensing of Born-Infeld BH~\cite{Jafarzade}. In this paper we will extend these results via studying the deflection of charged particles by a charged EGB BH using the Gauss-Bonnet (GB) theorem.

The geometric method of using the GB theorem to study the weak gravitational bending angle of light was first introduced by Gibbons and Werner~\cite{GW2008}. Later, Werner extended this method to stationary spacetime by using the technique of Randers-Finsler geometry~\cite{Werner2012}. The importance of Gibbons-Werner method is that it showed  the deflection angle can be connected to the global property of the geometry and has stimulated an impetuous growth of the studies of light deflection using GB theorem in recent years (see ~\cite{Jusufi:string17,Jusufi:monopole,Ali:BML, Ali:wormhole,Javed1,Javed2,Javed3,Moumni2020,Sakalli2017,Goulart2018,zhu2019,liucheng} and references therein). 
It inspired some authors to consider the finite distance gravitational deflection of light, a more general situation where the observer and source are both assumed to be at finite distance from a lens~\cite{ISOA2016,OIA2017,OA2019,Arakida2018,Arakida2020}.
Moreover, the Gibbons-Werner method has also been extended to the investigation of the deflection of NMPs which usually can not reach the speed of light either locally or asymptotically ~\cite{CG2018,CGJ2019,Jusufi:mp,LHZ2020}. In particular, the finite distance effect of NMP were considered via GB theorem~\cite{LJ2020,LA2020} and perturbative method~\cite{Duan2020,HuangJia2020}. 

Recently, Crisnejo et al.  pointed out the correspondance between the motion of charged particle in an external repulsive field and the motion of light in a dispersive medium. Thus they can apply the GB theorem to the corresponding optical metric to study the deflection of CMP in Reissner-Nordst\"{o}m (RN) spacetime~\cite{CGV:charged}. Later, Jusufi extended this work to Kerr-Newman spacetime~\cite{Jusufi:cmp}. In the present work, we will apply the GB theorem to investigate the deflection of 
CMP by a charged EGB BH under the influence of both electrostatic and gravitational interaction, and the finite distance effect will also be considered. Towards this purpose, we will use the Jacobi metric of a CMP. 

This paper is organized as follows. In Sec.~\ref{Jacotric}, we first review the Jacobi metric for a CMP in static spacetime, then we will use the Jacobi metric to drive the equation of motion of a CMP in an SSS spacetime. In Sec.~\ref{motion}, we obtain the Jacobi metric of charged EGB BH and further study the motion of a CMP in the weak field approximation. In Sec.~\ref{lensing}, we calculate the finite distance deflection of CMP using the GB theorem. Moreover, the effects of the BH charge, the signal charge-to-mass ratio, the coupling constant of the GB term, and the particle velocity on deflection angle are analyzed. Finally, we end our paper with a short conclusion in Sec.~\ref{conclusion}. In this paper, Greek and Latin indices are used to denote spacetime and spacial coordinates respectively. We use the natural units $G = c =4\pi \varepsilon_0= 1$ and the metric signature $(-,+,+,+)$.

\section{Jacobi metric for a charged particle in static and spherically symmetric spacetimes}\label{Jacotric}

The motion of a CMP of mass $m$ and charge $q$ in charged BH spacetime is described by the Lorentz equation~\cite{Rohrlich,Tursunov}
\begin{eqnarray}
\frac{d^2{x}^{\rho}}{d\tau^2}+\Gamma^{\rho}_{\mu\nu}\frac{dx^{\mu}}{d\tau}\frac{dx^{\nu}}{d\tau}= \frac{q}{m} F^{\rho}_{\mu}\frac{d{x}^{\mu}}{d\tau}~,
\end{eqnarray}
where $\tau$ is the proper time of the particle, and $F_{\mu\nu}=\partial_{\mu}A_{\nu}-\partial_{\nu}A_{\mu}$ is the field strength tensor with $A_{\mu}$ being the electromagnetic potential. From this we can see that the motion of the particle is no longer a geodesic in the background spacetime.

Jacobi metric is a basic tool in geometric dynamics, which is the subject of applying differential geometry to the dynamics of mechanical systems~\cite{Awrejcewicz}.  Gibbons first established the Jacobi metric for NMP in static spacetime~\cite{Gibbons2016}. Das and Ghosh extended this work to study the motion of CMP in RN spacetime~\cite{Das2017}. Chanda et al.  derived the Jacobi metric for a NMP in stationary spacetime~\cite{Chanda2019}. 
The Jacobi metrics related to the more general Lagrangian can be found in Refs.~\cite{Chanda1,Maraner}.

Let us review the Jacobi metric for a CMP in static spacetime. For a generic static metric,
\begin{eqnarray}
ds^2= g_{tt}dt^2+ g_{ij}dx^i dx^j~,
\end{eqnarray}
the mechanical action of a CMP can be written as
\begin{eqnarray}
\mathcal{S}&=&\int_{t1}^{t2}dt\mathcal{L}\nn\\
&=&\int_{t1}^{t2}\left(-m\sqrt{-{g}_{tt}-g_{ij}\dot{x}^{i}\dot{x}^{j}}+q A_{\mu}\dot{x}^{\mu}\right)dt,
\end{eqnarray}
where a dot denotes derivative with respect to $t$. We suppose that $A_{\mu}$ is an electrostatic potential, that is $A_{\mu}dx^{\mu}=A_0dt$ , then the Lagrangian becomes
\begin{eqnarray}
\mathcal{L}&=&-m\sqrt{-{g}_{tt}-g_{ij}\dot{x}^{i}\dot{x}^{j}}+qA_{0}.
\end{eqnarray}
From this, we can write the canonical momenta as follows,
\begin{eqnarray}
p_{i}=\frac{\partial \mathcal{L}}{\partial \dot{x}^i}=\frac{mg_{ij}\dot{x}^j}{\sqrt{-{g}_{tt}-g_{ij}\dot{x}^{i}\dot{x}^{j}}}.
\end{eqnarray}
The Jacobi metric thus can be written as~\cite{Chanda2019}
\begin{eqnarray}
ds_J=p_i\dot{x}^i=\frac{mg_{ij}\dot{x}^i\dot{x}^j}{\sqrt{-g_{tt}-g_{ij}\dot{x}^i\dot{x}^j}}=\sqrt{J_{ij}\dot{x}^i\dot{x}^j},
\end{eqnarray}
which satisfies
\begin{eqnarray}
\label{constraint}
J^{ij}p_ip_j=1.
\end{eqnarray}
In addition, the Hamiltonian is
\begin{eqnarray}
\mathcal{H}&=&p_{i}\dot{x}^{i}-\mathcal{L}\nn\\
&=&\frac{-mg_{tt}}{\sqrt{-{g}_{tt}-g_{ij}\dot{x}^{i}\dot{x}^{j}}}-qA_0\nn\\
&=&\sqrt{-g_{tt}m^2-g_{tt}g^{ij}p_ip_j}-qA_0=E,
\end{eqnarray}
with $E$ being the energy of the particle. The above equation leads to
\begin{eqnarray}
\frac{-g_{tt}g^{ij}p_ip_j}{\left(E+qA_0\right)^2+m^2g_{tt}}=1.
\end{eqnarray}
Comparing this equation with~\eqref{constraint}, we can see immediately
\begin{eqnarray}
J^{ij}=\frac{-g_{tt}g^{ij}}{\left(E+qA_0\right)^2+m^2g_{tt}}.
\end{eqnarray}
Then by $J_{ij}J^{ik}=\delta_j^k$, the Jacobi metric reads
\begin{eqnarray}
\label{jacobimetric}
J_{ij}&=&\left[\left(E+q A_0\right)^2+m^2{g}_{tt}\right]\frac{g_{ij}}{-{g}_{tt}}.
\end{eqnarray}

Eq.~\eqref{jacobimetric} is the starting point for the study of the deflection of CMP using GB theorem, and its importance relies on the fact that the motion of a CMP follows the geodesic in this metric. If there is no electric potential, i.e., $A_0=0$, Jacobi metric~\eqref{jacobimetric} reduces to the NMP case~\cite{Gibbons2016},
\begin{eqnarray}
\label{opticalmetric}
 &&J_{ij}=\frac{\left(E^2+m^2g_{tt}\right)}{-g_{tt}}g_{ij},
\end{eqnarray}
which has been used to investigate the gravitational deflection angle of NMP in
a static spacetime geometry~\cite{LHZ2020}. 

In the following, the Jacobi metric~\eqref{jacobimetric} will be used to derive the equation of motion of a CMP in SSS spacetime whose metric is given by
\begin{eqnarray}
\label{eq:sssmetric}
 ds^2=-A(r)dt^2+B(r)dr^2+C(r)(d\theta^2+\sin^2\theta d\phi^2).
\end{eqnarray}
By~\eqref{jacobimetric}, the Jacobi metric for a CMP in this background reads
\begin{eqnarray}
 dl^2&=&\left[(E+qA_0)^2-m^2A\right]\nn\\
 &&\times\left[\frac{B}{A}dr^2+\frac{C}{A}\left(d\theta^2+\sin^2\theta d\phi^2\right)\right]. \label{eq:sssmetricindj}
\end{eqnarray}
Without losing any generality, we assume that the particle moves in the equatorial plane, that is $\theta=\pi/2$, then Eq. \eqref{eq:sssmetricindj} becomes
\begin{eqnarray}
\label{JSSSm}
 dl^2=\left[(E+qA_0)^2-m^2A\right]\left(\frac{B}{A}dr^2+\frac{C}{A}d\phi^2\right).
\end{eqnarray}
According to spherical symmetry, the angular momentum of the motion will be a constant
\begin{eqnarray}
\label{LSSSm}
 L=\left[(E+qA_0)^2-m^2A\right]\frac{C}{A}\left(\frac{d\phi}{dl}\right)=constant.
\end{eqnarray}
Using a new variable $u=1/r$, and Eqs. \eqref{JSSSm} and~\eqref{LSSSm}, we can obtain the orbit equation of a CMP moving in equatorial plane in terms of $u(\phi)$ as follows,
\begin{eqnarray}
\label{orbit18}
 \left(\frac{du}{d\phi}\right)^2=\frac{C^2u^4}{ABL^2}\left[\left(E+qA_0\right)^2-A\left(m^2+\frac{L^2}{C}\right)\right],~~~
\end{eqnarray}
In addition, the energy and angular momentum of the particle at infinity for an asymptotic observer satisfy
\begin{eqnarray}
\label{energy}
 E=\frac{m}{\sqrt{1-v^2}},~~~~~L=\frac{mvb}{\sqrt{1-v^2}},
\end{eqnarray}
where $v$ is the asymptotic velocity of the particle, and $b$ is the impact parameter. Clearly, they have a relation
\begin{eqnarray}
 b=\frac{L}{vE}.
\end{eqnarray}
Using~\eqref{energy}, we can rewrite the Jacobi metric~\eqref{JSSSm} as
\begin{align}
\label{Jacobimetric}
dl^2=&J_{ij}dx^idx^j\nn\\
 =&m^2\left[\left(\frac{1}{\sqrt{1-v^2}}+\frac{qA_0}{m}\right)^2-A\right]\left(\frac{B}{A}dr^2+\frac{C}{A}d\phi^2\right),
\end{align}
and the trajectory of the CMP~\eqref{orbit18} as
\begin{eqnarray}
\label{particleorbit}
 \left(\frac{du}{d\phi}\right)^2&=&\frac{C^2u^4}{ABv^2b^2}\bigg[\left(1+\frac{\sqrt{1-v^2}qA_0}{m}\right)^2\nn\\
 &&-A\left(1-v^2+\frac{b^2v^2}{C}\right)\bigg].~~~
\end{eqnarray}
\section{The motion of charged particle in the weak field limit}\label{motion}
The action of the Einstein-Maxwell-Gauss-Bonnet theory in a $D$-dimensional spacetime reads~\cite{Fernandes2020}
\begin{eqnarray}
\label{action}
 S=\frac{1}{16\pi}\int d^Dx \sqrt{-g}\left[R+\frac{\alpha}{D-4} \mathcal{G}-F_{\mu\nu}F^{\mu\nu}\right],
\end{eqnarray}
where $g$ is the determinant of the spacetime metric, $\alpha$ is the GB term coupling constant, and $\mathcal{G}$ is the GB term, defined by
\begin{eqnarray}
 \mathcal{G}=R^2-4R_{\mu\nu}R^{\mu\nu}+R_{\mu\nu\rho\sigma}R^{\mu\nu\rho\sigma}.
\end{eqnarray}
Here, $R$, $R_{\mu\nu}$ and $R_{\mu\nu\rho\sigma}$ are the Ricci scalar, Ricci tensor and Riemann tensor, respectively.

Note that in action~\eqref{action} we have rescaled the coupling constant $\alpha$ to $\alpha/(D-4)$. Considering the limit $D\to 4$, the charged EGB BH solution of action~\eqref{action} was obtained as~\cite{Fernandes2020}
\begin{eqnarray}
\label{thesolution}
 ds^2=-f(r)dt^2+\frac{1}{f(r)}dr^2+r^2\left(d\theta^2+\sin^2\theta d\phi^2\right),
 \end{eqnarray}
 where $f(r)$ and the vector potential are
 \begin{eqnarray}
 &&f(r)=1+\frac{r^2}{2\alpha}\left[1-\sqrt{1+4\alpha \left(\frac{2M}{r^3}-\frac{Q^2}{r^4}\right)}\right],\nn\\
  &&A_{\mu}dx^{\mu}=-\frac{Q}{r}dt,\label{eq:thevecpot}
\end{eqnarray}
with $M$ and $Q$ being the mass and charge of the BH, respectively. 
The solution~\eqref{thesolution} reduces to the RN spacetime if we take the limit $\alpha\to 0$. In addition, Eq.~\eqref{thesolution} has the same form as solutions in the gravity with a conformal anomaly~\cite{cai2010}, and in the gravity with quantum corrections~\cite{Tomozawa,Cognola}. Notice that the GB coupling constant $\alpha$ could take a negative value~\cite{GuoLi2020,ZhangCY}. 

For charged BH~\eqref{thesolution}, we have
\begin{eqnarray}
\label{metricfunction}
&& A(r)=B(r)^{-1}=f(r), ~C(r)=r^2, \nn\\
 &&A_0=-\frac{Q}{r}.
\end{eqnarray}
Substituting them into Eq.~\eqref{Jacobimetric}, we can write the two-dimensional Jacobi metric as follows,
\begin{eqnarray}
\label{jacobim}
 dl^2&=&J_{ij}dx^idx^j\nn\\
 &=&m^2\left[\left(\frac{1}{\sqrt{1-v^2}}-\frac{q Q}{mr}\right)^2-f(r)\right]\nn\\
 &&\times\left[\frac{1}{f^2(r)}dr^2+\frac{r^2 }{f(r)}d\phi^2\right],
\end{eqnarray}
In this paper we will only be interested in the weak field deflection, and therefore we can expand the metric components of \eqref{jacobim} in the large $r$ limit. In this way, the non-zero components of Jacobi metric are given by the following expressions,
\begin{eqnarray}
\label{EGBjacobi1}
 J_{rr}&=&\frac{m^2\left( 1+v^2 \right)}{1-v^2}\bigg[ \frac{v^2}{1+v^2}+\frac{2M}{r}-\frac{Q^2}{r^2}\nn\\
 &&+\frac{4\left( 2+v^2 \right)M^2}{\left(1+v^2\right)r^2}-\frac{4\alpha M^2}{r^4} \bigg] \nn\\
 &&-\frac{2mq}{\sqrt{1-v^2}}\left( \frac{Q}{r}+\frac{4MQ}{r^2} \right)+\frac{q^2Q^2}{r^2}\nn\\
 &&+\mathcal{O}\left(M^3,Q^3,MQ^2,M^2Q,\alpha^2\right),\\
 \label{EGBjacobi2}
 J_{\phi\phi}&=&\frac{m^2}{1-v^2}\bigg( r^2v^2+2Mr+4M^2-Q^2\nn\\
 &&-\frac{4\alpha M^2}{r^2} \bigg)-\frac{2mq\left( Qr+2MQ \right)}{\sqrt{1-v^2}}\nn\\
 &&+q^2Q^2+\mathcal{O}\left(M^3,Q^3,MQ^2,M^2Q,\alpha^2\right).
\end{eqnarray}
We can see that up to the above orders, the coupling constant $\alpha$ appears in the $M^2\alpha$ order and the  $Q^2\alpha$ term dose not show up.

Substituting~\eqref{metricfunction} into~\eqref{particleorbit}, the orbit equation becomes
\begin{eqnarray}
 \left(\frac{du}{d\phi}\right)^2&=&\left(\frac{1}{vb}-\frac{\sqrt{1-v^2}qQu}{mbv}\right)^2\nn\\
 &&-f(u)\left(\frac{1-v^2}{b^2v^2}+u^2\right),~~~
\end{eqnarray}
Usually, solving this equation is difficult. Again, since we are interested in the weak field limit, the problem can be greatly simplified. Using the condition $\frac{du}{d\phi}\mid_{\phi=\frac{\pi}{2}}=0$, one can use the iterative method (See Ref.~\cite{Arakida2012} for details) to obtain the following solution
\begin{eqnarray}
\label{orbitequation}
u&=&\frac{\sin \phi }{b}+\frac{1+v^2\cos ^2 \phi }{v^2}\frac{M}{b^2}\nn\\
&&-\frac{\sqrt{1-v^2}}{mv^2}\frac{qQ}{b^2}+\mathcal{O}\left(M^2,Q^2\right).
\end{eqnarray}
This solution then can be inverted perturbatively using the Lagrange Inversion theorem to express $\phi$ in terms of $u$. The result is found to be 
\begin{eqnarray}
\label{angle}
\phi(u)\approx\begin{cases}
\arcsin{(bu)}-M\phi_1+qQ\phi_2,    &\text{if } \vert{\phi}\vert <\frac{\pi}{2};~~\\
\pi-\arcsin{(bu)}+M\phi_1-qQ\phi_2,&\text{if } \vert {\phi}\vert >\frac{\pi}{2},~~
\end{cases} \label{eq:phitwosol}
\end{eqnarray}
where
\begin{eqnarray}
&&\phi_1=\frac{1+v^2-b^2u^2v^2}{b^2\sqrt{1-b^2u^2}v^2}~,~\phi_2=\frac{\sqrt{1-v^2}}{bm\sqrt{1-b^2u^2}v^2}~.\nn
\end{eqnarray}
\section{Deflection angle of charged massive particle}\label{lensing}

\subsection{Gauss-Bonnet theorem and lens geometry}
 Let $\mathcal{D}_a$ be a subset of a compact, oriented surface, with Gaussian curvature $\mathcal{K}$ and Euler characteristic number $\chi(\mathcal{D}_a)$, and its boundary $\partial{\mathcal{D}_a}$ a piecewise smooth curve with geodesic curvature $k$, the GB formula reads~\cite{Carmo1976}
\begin{equation}
\label{GBT}
\iint_{\mathcal{D}_a}{\mathcal{K}}dS+\oint_{\partial \mathcal{D}_a} k~d\sigma+\sum_{i=1}{\beta_i}=2\pi\chi(\mathcal{D}_a),\\
\end{equation}
where $dS$ is the area element, $d\sigma$ is the line element of boundary, and $\beta_i$ is the jump angle in the $i$-th vertex of $\partial{\mathcal{D}_a}$ in the positive sense. The three terms on the left side of Eq.~\eqref{GBT} correspond to surface curvature, line curvature and point curvature, respectively, and the right side is the Euler characteristic number. In other words, the GB theorem reveals the relation between the curvature of a Riemannian metric and the topology of the manifold.

\begin{figure}[htp]
\centering
\includegraphics[width=8.0cm]{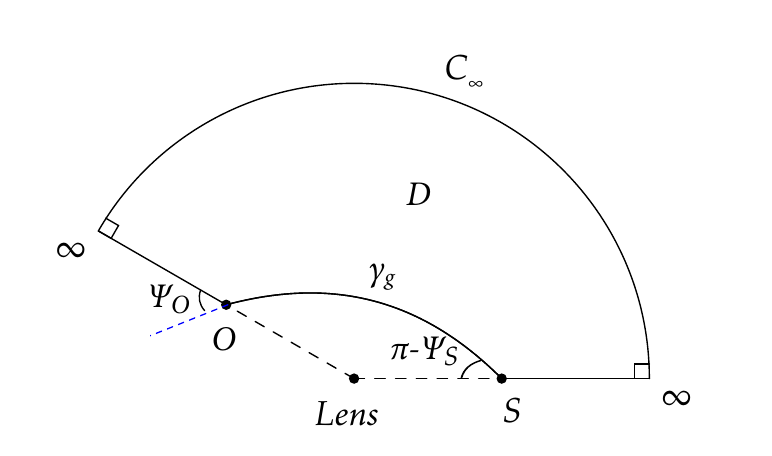}
\caption{A region $\mathcal{D}\subset(\mathcal{M},J_{ij})$. Notice that $\beta_S=\pi-\Psi_S$ and $\beta_O=\Psi_O$. The radial and angular coordinates of $O$ and $S$ are $(r_O,\phi_O)$ and $(r_S,\phi_S)$. }
\label{figure}
\end{figure}

In the Jacobi space $(\mathcal{M},J_{ij})$, a CMP follows the geodesic from source ($S$) to observer ($O$) and is deflected by a lens (see Fig.~\ref{figure} for the illustration). In the following, the method of Ref.~\cite{ISOA2016} will be used to study the deflection angle $\delta$, which is defined as
\begin{eqnarray}
\label{angledef}
&& \delta\equiv \Psi_O-\Psi_S+\phi_{OS},
\end{eqnarray}
where $\Psi_O$ and $\Psi_S$ are angles between the tangent of the particle ray and the radial direction from the lens to observer and source, respectively, and the change of the angular coordinate $\phi_{OS}\equiv\phi_O-\phi_S$.

Let us choose $\mathcal{D}_a$ to be the region $\mathcal{D} \subset(\mathcal{M},J_{ij})$ bounded by four curves, as shown in Fig. \ref{figure}. Three of these four curves are geodesics, including the particle ray $\gamma_{g}$ and two spatial geodesics of outgoing radial lines passing through $O$ and $S$ respectively, and one is a non-geodesic circular arc segment $C_{\infty}$, i.e., $C_{r}$ in the $r\to\infty$ limit. Notice that only $C_{\infty}$ has a non-zero geodesic curvature, and two jump angles where $C_{\infty}$ and the radial curves intersect are both $\pi/2$. In addition, Euler characteristic number $\chi(\mathcal{D})=1$ because $\mathcal{D}$ is a non-singular region.

Applying GB theorem~\eqref{GBT} to region $\mathcal{D}$, we have
\begin{eqnarray}
\label{GBGBGB}
\iint_{D}\mathcal{K} dS+\int_{\phi_S}^{\phi_O} \left[\left(k \frac{dl}{d\phi}\right)\left({C_{\infty}}\right)\right] d\phi +\beta_S+\beta_O=\pi.~~~
\end{eqnarray}
Here, we have (see Eq. (23) of Ref.~\cite{LHZ2020})
\begin{eqnarray}
\left(k \frac{dl}{d\phi}\right)\left({C_{\infty}}\right)=\lim_{r_c\to  \infty}\sqrt{\frac{B(r_c)}{C(r_c)}\left(\bar{\Gamma}^r_{\phi\phi}(r_c)\right)^2},
\end{eqnarray}
with $\bar{\Gamma}^k_{ij}$ being the Christoffel symbols of $J_{ij}$.
Using Eqs.~\eqref{metricfunction} and~\eqref{jacobim}, we find $\left(k \frac{dl}{d\phi}\right)\left({C_{\infty}}\right)=1$, which shows that our Jacobi metric is asymptotically Euclidean. Then considering $\beta_O=\Psi_O$ and $\beta_S=\pi-\Psi_S$, and using the definition~\eqref{angledef}, Eq.~\eqref{GBGBGB} leads to
\begin{eqnarray}
\label{Gauss-Bonnet}
\delta=-\iint_{\mathcal{D}}\mathcal{K} dS.
\end{eqnarray}
This two dimensional integral can be expressed explicitly as
\begin{eqnarray}
\label{GaussBonnet}
\delta=-\int_{\phi_S}^{\phi_O}\int_{r(\phi)=1/u(\phi)}^{\infty}\mathcal{K}\sqrt{J}drd\phi,
\end{eqnarray}
where $J$ is the determinant of $J_{ij}$.
The Gaussian curvature $\mathcal{K}$ can be calculated by~\cite{Werner2012}
\begin{eqnarray}
\label{Gauss-K}
\mathcal{K}&=&\frac{1}{\sqrt{J}}\left[\frac{\partial }{\partial{\phi}}\left(\frac{\sqrt{J}}{J_{rr}}{\bar{\Gamma}^\phi_{rr}}\right)-\frac{\partial}{\partial{r}}\left(\frac{\sqrt{J }}{J_{rr}}{\bar{\Gamma}^\phi_{r\phi}}\right)\right].~~~~
\end{eqnarray}
Therefore the task of finding the deflection angle in a SSS spacetime reduces to the integration of Eq. \eqref{GaussBonnet}.

\subsection{Deflection angle of CMP in the charged EGB BH}

While the above procedure is general, i.e., applicable to arbitrary SSS spacetimes, to proceed we will have to go to specific spacetime, in this case the EGB BH described by Eqs. \eqref{thesolution} and \eqref{eq:thevecpot}. 
We first calculate the Gaussian curvature $\mathcal{K}$ in Eq. \eqref{GaussBonnet}. Substituting \eqref{EGBjacobi1} and~\eqref{EGBjacobi2} into~\eqref{Gauss-K}, the result is found to be
\begin{eqnarray}
\label{GaussCurvature}
 \mathcal{K}&=&\frac{\left( 1-v^2 \right)}{m^2v^4}\bigg[ -\frac{\left( 1+v^2 \right) M}{r^3}+\frac{3\left( 2-v^2 \right) M^2}{r^4v^2}\nn\\
 &&+\frac{8\left( 4+v^2 \right) \left( M^2\alpha \right)}{r^6}+\frac{\left( 2+v^2 \right) Q^2}{r^4} \bigg]\nn\\
  &&+\frac{qQ\left( 1-v^2 \right) ^{\text{3/}2}}{m^3v^6}\left[ \frac{v^2}{r^3}-\frac{3M\left( 4-v^2 \right)}{r^4} \right] \nn\\
  &&+\frac{2\left( 3-v^2 \right) \left( 1-v^2 \right) ^2}{m^4v^6}\frac{q^2Q^2}{r^4}\nn\\
  &&+\mathcal{O}\left(M^3,Q^3,MQ^2,M^2Q,\alpha^2\right).
\end{eqnarray}
Secondly, the integral limits of Eq. \eqref{GaussBonnet} should be clarified. Our lensing setup in Fig. \ref{figure} supposed that $\phi_O>\frac{\pi}{2}>\phi_S$, and then according to Eq.~\eqref{angle} the coordinate angle of the source and the observer becomes,
\begin{eqnarray}
\label{coordinateangleS}
\phi_S&=&\arcsin \left( bu_S \right) -\left(\Delta u_S+\frac{1}{v^2\Delta u_S}\right)\frac{M}{b}\nn\\
&&+\frac{\sqrt{1-v^2}}{mv^2\Delta u_S}\frac{qQ}{b}+\mathcal{O}\left(M^2,Q^2\right),\\
\label{coordinateangleR}
\phi_O&=&\pi -\arcsin \left( bu_O \right) +\left(\Delta u_O+\frac{1}{v^2\Delta u_O}\right)\frac{M}{b}\nn\\
&&-\frac{\sqrt{1-v^2}}{mv^2\Delta u_O}\frac{qQ}{b}+\mathcal{O}\left(M^2,Q^2\right),
\end{eqnarray}
where $u_i=1/r_i~(i=O,S)$ and $\Delta u_i=\sqrt{1-b^2u_i^2}$.
After obtaining Gaussian curvature~\eqref{GaussCurvature}, particle orbit ~\eqref{orbitequation}, and coordinate angles~\eqref{coordinateangleS} and~\eqref{coordinateangleR}, now we can calculate the finite distance deflection angle of CMP by a four-dimensional charged EGB BH according to Eq.~\eqref{GaussBonnet}. Fortunately, the integral is only tedious but not difficult. The result is given by
\begin{eqnarray}
\label{Sfangle}
\delta&=&\delta_1\frac{M}{b}+\delta_2\frac{q_mQ}{b}+\delta_3\frac{Q^2}{b^2}+\delta_4\frac{q_m^2Q^2}{b^2}\nn\\
&&+\delta_5\frac{q_mQM}{b^2}+\delta_6\frac{M^2}{b^2}+\delta_7\frac{M^2\alpha}{b^4}\nn\\
&&+\mathcal{O}\left(M^3,Q^3,MQ^2,M^2Q,\alpha^2\right),
\end{eqnarray}
 where
\begin{eqnarray}
\delta_1&=&\left(\Delta \text{u}_O+\Delta \text{u}_S\right)\left( 1+\frac{1}{v^2} \right),\nn\\
\delta_2&=&-\left(\Delta \text{u}_O+\Delta \text{u}_S\right)\frac{\sqrt{1-v^2}}{v^2},\nn\\
\delta_3&=&-\left( \Xi+\Gamma \right)\left( \frac{1}{4}+\frac{1}{2v^2}\right),\nn\\
\delta_4&=&\frac{\left( 1-v^2 \right) }{2v^4}\bigg[ \Xi v^2+2b\Omega-\left( 2-v^2 \right) \Gamma\bigg],\nn\\
\delta_5&=&\frac{\sqrt{1-v^2}}{v^2}\bigg[ -3\Xi -\frac{2\left( 1+v^2 \right)}{v^2}\Omega\nn\\
&&+ \frac{2-v^2}{v^2}\Gamma+\frac{b^3u_{O}^{3}}{\Delta \text{u}_O}+\frac{b^3u_{S}^{3}}{\Delta \text{u}_S} \bigg],\nn\\
\delta_6&=&\frac{ 3}{4}\left( 1+\frac{4}{v^2} \right) \left( \Xi +\Omega \right) \nn\\
&&+\frac{b^3\left( 4-8v^2-3v^4 \right) }{4v^4}\left(\frac{u_{O}^{3}}{\Delta \text{u}_O}+\frac{u_{S}^{3}}{\Delta \text{u}_S}\right) ,\nn\\
\delta_7&=&-\frac{\left( 4+v^2 \right)}{4v^2} \bigg[ 3\Xi +4\Gamma+b^3\left(u_O^3\Delta \text{u}_O+u_S^3\Delta \text{u}_S\right)\nn\\
&&-bu_O\left(\Delta \text{u}_{O}\right)^{3} -bu_S\left(\Delta \text{u}_{S}\right)^{3} \bigg]\nn,
\end{eqnarray}
with 
\begin{eqnarray}
&&q_m=q/m,\nn\\
&&\Xi=\pi-\arcsin(bu_O)-\arcsin(bu_S),\nn\\
&&\Gamma=b\left(u_O\Delta \text{u}_O+u_S\Delta \text{u}_S\right),\nn\\
&&\Omega=b\left(\frac{u_O}{\Delta \text{u}_O}+\frac{u_S}{\Delta \text{u}_S}\right),\nn\\
&& \Delta u_i=\sqrt{1-b^2u_i^2}~(i=O,S).\nn
\end{eqnarray}

We emphasis that the Eq. \eqref{Sfangle} is a very comprehensive result because it contains dependence on both the spacetime parameters, such as $M,~Q,~\alpha$, and signal parameters such as $v,~q,~m$. And moreover, the effects of geometrical parameters $r_S=1/u_S$ and $r_O=1/u_O$ are also included. Therefore there is a quite large parameter space  we can study, and a few limits we can take to compare with previously known results in simpler cases.

\begin{figure}[htp]
\includegraphics[width=0.4\textwidth]{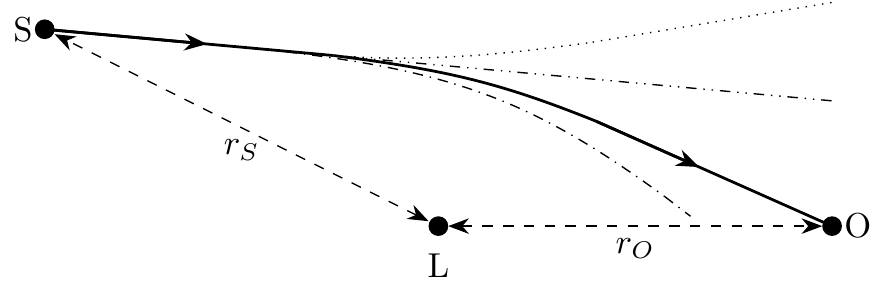}
\caption{The deflection of charged signal by the charged lens $L$ from source $S$ to observer $O$. The gravitational interaction always attracts the signal particle (dot-dashed line) while the electrostatic interaction might attract or repel (dotted line) the CMP depending on the sign of $Q$ and $q$. The true trajectory is the combined one plotted using solid curve.}
\label{fig:lenselec}
\end{figure}

Among the spacetime parameters, $M$ provides a basic length scale against which all other quantities with the same dimension, i.e., $Q,~\sqrt{\alpha},~q,~m,~r_S,~r_O$ can be compared with. Therefore it can be interpreted as the main (although not the sole) source of gravity in this spacetime. $Q$ on the other hand, provides not only extra gravitational effect, but also a direct source to the electrostatic interaction with the charged signal (see Fig. \ref{fig:lenselec}). Finally the GB coupling $\alpha$ controls the amount of deviation of the spacetime from the RN solution. Now for the geometrical parameters $r_S,~r_O$ as well as the signal parameter $v$, they are actually determined by the initial and final conditions and therefore quite free. For the other signal parameters $m$ and $q$, due to general equivalence principle, $m$ will not appear in the deflection if there was only gravity, which is the case if $q=0$. In other words, $q/m$ always appears in this fixed form in the deflection angle.  

Observing the above, we find there are the following limits one can take to compare with known literature. In the RN and infinite source/observer distance limits, setting $\alpha=0$ and $r_S=r_O=\infty$, Eq. \eqref{Sfangle} reduces to Eq. (183) of Ref. \cite{CGV:charged} or Eq. (48) of Ref. \cite{Jusufi:cmp}. On the other hand, taking the RN and NMP limits by letting $\alpha =0$ and $q_m= 0$, Eq. \eqref{Sfangle} agrees with Eq. (5.1) of Ref. \cite{HuangJia2020} after setting its spacetime spin $a$ to zero. If one further sets $r_S,~r_O$ to infinity, then the Eq. (52) in Ref. \cite{Jiaepjc2020} is recovered. Moreover, if we set $Q=\alpha=0$ in Eq.  \eqref{Sfangle} but keep $r_S,~r_O$ finite, then the charge of the signal would automatically become ineffective to the deflection angle, and Eq. (42) of Ref. \cite{LJ2020} after setting its spacetime spin $a$ to zero, and Eq. (B4) of Ref. \cite{HuangJia2020} after setting high order metric coefficients to zero, are obtained. 

Now we return back to some limits in the case that GB coupling constant $\alpha$ is nonzero.  If we consider the asymptotic case that both the observer and source are at infinite distance, then letting $r_S=r_O=\infty$, result \eqref{Sfangle} leads to the asymptotic deflection angle,
\begin{eqnarray}
\label{iangle}
 \delta_{\infty}
 &=&
\left( 1+\frac{1}{v^2} \right) \frac{2M}{b}-\frac{\sqrt{1-v^2}}{v^2}\frac{2q_mQ}{b}\nn\\
&&+\left( 1+\frac{4}{v^2} \right) \frac{3\pi M^2}{4b^2}-\left( 1+\frac{2}{v^2} \right) \frac{\pi Q^2}{4b^2}\nn\\
&&-\frac{\sqrt{1-v^2}}{v^2}\frac{3\pi q_mMQ}{b^2}+\frac{1-v^2}{v^2}\frac{\pi q_m^2Q^2}{2b^2}\nn\\
&&-\left( 1+\frac{4}{v^2} \right) \frac{3\pi M^2\alpha}{4b^4}\nn\\
&&+\mathcal{O}\left(M^3,Q^3,MQ^2,M^2Q,\alpha^2\right).
\end{eqnarray}
Furthermore, for a NMP, setting $q_m=0$, the above expression becomes
\begin{eqnarray}
\label{infiniteangle:np}
 \delta_{\infty,g}
 &=&
\left( 1+\frac{1}{v^2} \right) \frac{2M}{b}-\left( 1+\frac{2}{v^2} \right) \frac{\pi Q^2}{4b^2}\nn\\
&&+\left( 1+\frac{4}{v^2} \right) \frac{3\pi M^2}{4b^2}\left(1-\frac{\alpha}{b^2}\right)\nn\\
&&+\mathcal{O}\left(M^3,Q^3,MQ^2,M^2Q,\alpha^2\right).
\end{eqnarray}
On the other hand, if the signal is null and neutral like light or GW, then further setting $v=1$ in Eq. \eqref{infiniteangle:np} yields
\begin{eqnarray}
 \delta_{\infty,g,null}&=&\frac{4M}{b}-\frac{3\pi Q^2}{4b^2}+\frac{15\pi M^2}{4b^2}\left(1-\frac{\alpha}{b^2}\right)\nn\\
 &&+\mathcal{O}\left(M^3,Q^3,MQ^2,M^2Q,\alpha^2\right)~.
\end{eqnarray}
When we neglect the terms containing charge $Q$, this expression is consistent with Eq. (18) of Ref.~\cite{Heydari-Fard}.

\subsection{Gauss-Bonnet theorem method applied to motion in pure electrostatic field}

In the above limits of the deflection angle, we largely focused on the effect of gravitational interaction.  However, since there are two kinds of interactions in this situation, it would also be very inspiring to consider the electrostatic limit of the deflection. This limit can be approached by setting $M\to 0$ in Eq. \eqref{Sfangle} and $Q\to 0$ in some but not all terms of it. The point is that, the charge $Q$ is responsible for both gravitational (although only partially) and electrostatic interactions, through respectively the $Q^2/r^2$ term in the metric and $-Q/r$ term in the Coulomb potential $A_0$, and we would like to turn off the gravitational part only and keep the Coulomb interaction. Therefore, we see that in Eq.  \eqref{Sfangle}, the terms containing $\delta_1,~\delta_3,~\delta_6$ and $\delta_7$ are of gravitational origin and can be collectively denoted as $\delta_g$, i.e., 
\begin{align}
    \delta_g=\delta_1\frac{M}{b}+\delta_3\frac{Q^2}{b^2}+\delta_6\frac{M^2}{b^2}+\delta_7\frac{M^2\alpha}{b^4}. 
\label{eq:dgcdef}
\end{align}
While the terms including $\delta_2$ and $\delta_4$ are due to electrostatic interaction and they are denoted as
\begin{eqnarray}
\delta_c&=&-\frac{\left(\Delta u_O+\Delta u_S\right)\sqrt{1-v^2}}{v^2}\frac{q_mQ}{b}\nn\\
&&+\left[ \bigg(\pi-\arcsin(bu_O)-\arcsin(bu_S)
\right)v^2\nn\\
&&-\left( 2-v^2 \right) b\left(u_O\Delta \text{u}_O+u_S\Delta \text{u}_S\right) \nn\\
&&+\frac{2bu_O}{\Delta \text{u}_O}+\frac{2bu_S}{\Delta \text{u}_S}\bigg]\frac{\left( 1-v^2 \right) }{2v^4}\frac{q_m^2Q^2}{b^2}+\mathcal{O}\left(Q^3\right). \label{eq:deltacoul}
\end{eqnarray}
Lastly, the term involving $\delta_5$ in Eq. \eqref{Sfangle} is proportional to both $q_m$ and $M$ and therefore represents a gravitational-electrostatic coupling, and it will be denoted as $\delta_{gc}$. This interaction is repulsive (or attractive) when $q_m$ and $Q$ has the same (or oppositve) signs.

In this subsection, we would like to show that the method using GB theorem with the induced Jacobi metric to calculate the deflection angle is applicable to the deflection caused purely by electrostatic potential, and the result agrees with Eq. \eqref{eq:deltacoul}.

When considering only the electrostatic interaction, we should start from the flat Minkowski metric in spherical coordinates 
\begin{eqnarray}
 ds^2=-dt^2+dr^2+r^2\left(d\theta^2+\sin^2\theta d\phi^2\right),
\end{eqnarray}
and the Coulomb potential field
\begin{eqnarray}
 A_\mu dx^\mu=-\frac{Q}{r}dt.
\end{eqnarray}
For this case, then the corresponding induced Jacobi metric, Eq. \eqref{Jacobimetric}, is found to be
\begin{eqnarray}
 dl^2=m^2\left[\left(\frac{1}{\sqrt{1-v^2}}-\frac{q Q}{mr}\right)^2-1\right]\left[dr^2+r^2 d\phi^2\right].
\end{eqnarray}
and the Gaussian curvature \eqref{Gauss-K} becomes
 \begin{equation}
 \mathcal{K}=\frac{\left( 1-v^2 \right) ^{\text{3/}2}}{m^3v^4} \frac{qQ}{r^3}+\frac{2\left( 3-v^2 \right) \left( 1-v^2 \right) ^2}{m^4v^6}\frac{q^2Q^2}{r^4}+\mathcal{O}(Q^3).
\end{equation}
Clearly, although we are studying problems in a flat spacetime, the Jacobi space of the particle is curved.

The equation of motion of the charged particle, Eq. \eqref{particleorbit}, becomes
\begin{eqnarray}
 \left(\frac{du}{d\phi}\right)^2=\left(\frac{1}{vb}-\frac{\sqrt{1-v^2}q Qu}{mvb}\right)^2-\frac{1-v^2}{b^2v^2}-u^2.~~~~~
\end{eqnarray}
From this, we can obtain to the leading order of $q$ and $Q$
\begin{eqnarray}
&&u=\frac{\sin \phi }{b}-\frac{\sqrt{1-v^2}}{mv^2}\frac{qQ}{b^2}+\mathcal{O}\left(Q^2\right),\\
&&\phi_S=\arcsin \left( bu_S \right) +\frac{\sqrt{1-v^2}}{v^2\Delta u_S}\frac{qQ}{mb}+\mathcal{O}\left(Q^2\right),\\
&&\phi_O=\pi -\arcsin \left( bu_O \right)-\frac{\sqrt{1-v^2}}{v^2\Delta u_O}\frac{qQ}{mb}+\mathcal{O}\left(Q^2\right).
\end{eqnarray}
Finally, applying the GB theorem to the deflection angle, i.e. Eq.~\eqref{GaussBonnet}, and carrying out the integral, the deflection of a charged particle in Coulomb potential is obtained as
\begin{eqnarray}
\delta_c&=&-\frac{\left(\Delta u_O+\Delta u_S\right)\sqrt{1-v^2}}{v^2}\frac{q_mQ}{b}\nn\\
&&+\left[ \bigg(\pi-\arcsin(bu_O)-\arcsin(bu_S)
\right)v^2\nn\\
&&-\left( 2-v^2 \right) b\left(u_O\Delta \text{u}_O+u_S\Delta \text{u}_S\right) \nn\\
&&+\frac{2bu_O}{\Delta \text{u}_O}+\frac{2bu_S}{\Delta \text{u}_S}\bigg]\frac{\left( 1-v^2 \right) }{2v^4}\frac{q_m^2Q^2}{b^2}+\mathcal{O}\left(Q^3\right). \label{eq:csfinite}
\end{eqnarray}
This agrees with Eq. \eqref{eq:deltacoul} perfectly.
The infinite distance limit, $u_O\to 0$ and $u_S\to 0$, of Eq. \eqref{eq:csfinite} then becomes
\begin{eqnarray}
\delta_{\infty,c}=-\frac{\sqrt{1-v^2}}{v^2}\frac{2q_mQ}{b}+\frac{1-v^2}{v^2}\frac{\pi q_m^2Q^2}{2b^2}+\mathcal{O}\left(Q^3\right). \label{eq:2condcol}
\end{eqnarray}
The first order term of this equation was given in Ref. \cite{Padmanabhan} for classical Coulomb scattering in the relativistic case. Its non-relativistic limit $v\to 0$ is the well-known Rutherford scattering formula. In general, Eq. \eqref{eq:csfinite} provides for Coulomb scattering a deflection angle formula with finite distance correction. 

\subsection{Effects of charges, parameter $\alpha$ and velocity $v$}

In astronomy, it is often assumed implicitly and by default that all astrophysical objects including compact objects (stars, BH, etc) and galaxies or their clusters are electrically neutral because of the possible selective accretion of the opposite charges from the surrounding environment. 
However, this is indeed an over simplification. It is known that astrophysical bodies should be slightly positively charged in order for electrons and protons in the stellar atmosphere to maintain quasi-local equilibrium \cite{eddington,bally}. This will result in a BH charge $Q_{eq}$ proportional to the BH mass with a coefficient about 100 [C] per sun mass, i.e.,
\begin{align}
    Q_{eq}\approx 100 \frac{M}{M_\odot}~[\text{C}]. 
\end{align}
If the charge of the BH is induced by magnetic field $B_{mag}$  around it, then the corresponding charge $Q_{mag}$ can be much larger \cite{Zajacek:2018vsj}
\begin{align}
    Q_{mag}\approx 1.46\times 10^2\left(\frac{M}{M_\odot}\right)^2\frac{B_{mag}}{10~[\text{G}]}~[\text{C}],
\end{align}
where $B_{mag}$ is typically of order 10 [G]. 
Of course, both these two charges are still very far from the extreme RN spacetime limit
\begin{align}
    Q_{ext}\approx 1.72\times 10^{20} \frac{M}{M_\odot}~[\text{C}]
\end{align}
for typical supermassive BHs (SMBH) in galaxy centers with $M\sim \mathcal{O}(10^6M_\odot)$. Observing all these, therefore we will limit the charge $Q$ in our analysis of the effect of charge to the deflection angle to a somewhat loosely fixed value $Q_l=\sqrt{Q_{mag}Q_{ext}}\approx 5.4\times 10^5 Q_{mag}=Q_{ext}/(5.4\times 10^5)$ .

\begin{figure}[htp]
\includegraphics[width=0.4\textwidth]{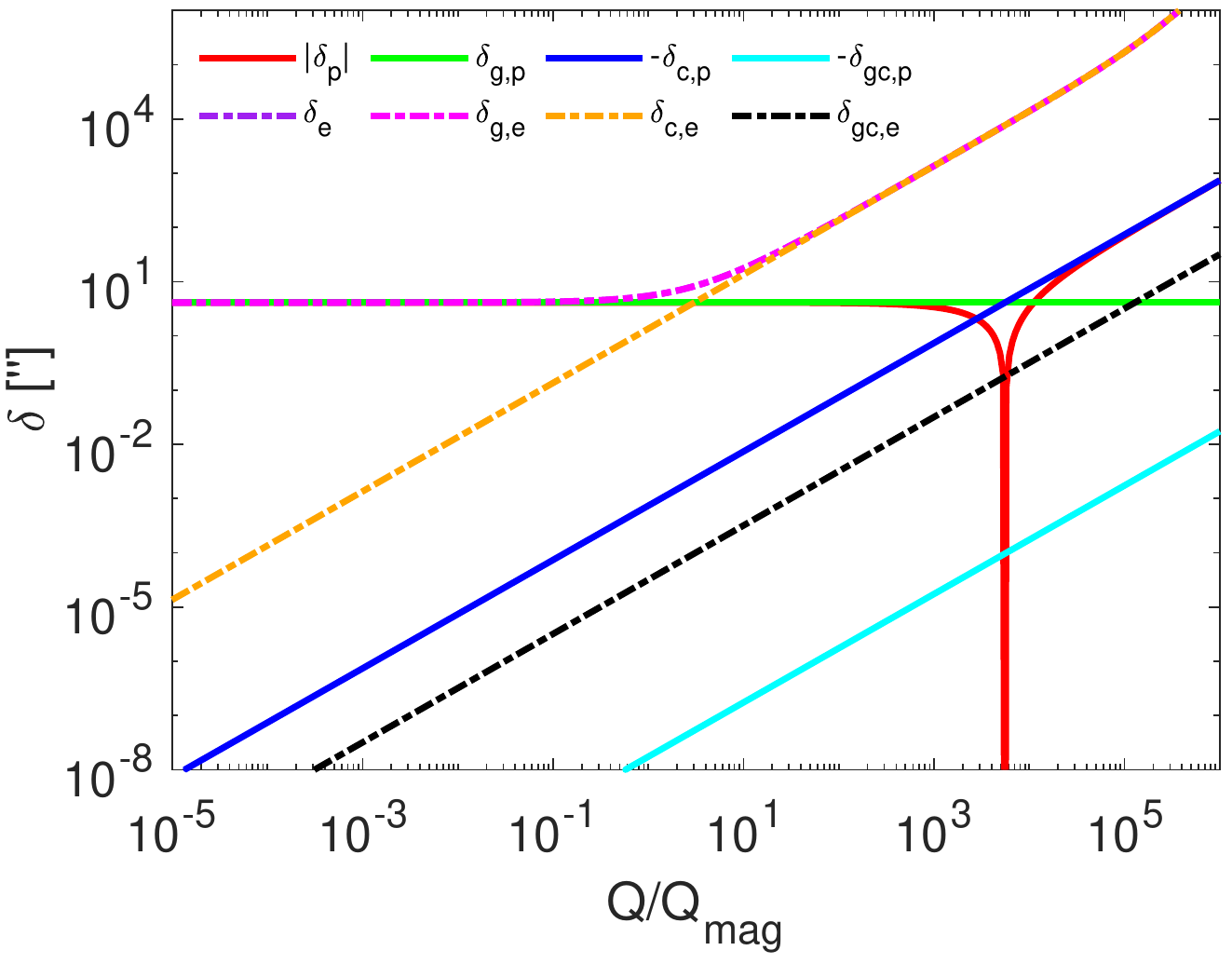}\\
(a)\\
\includegraphics[width=0.4\textwidth]{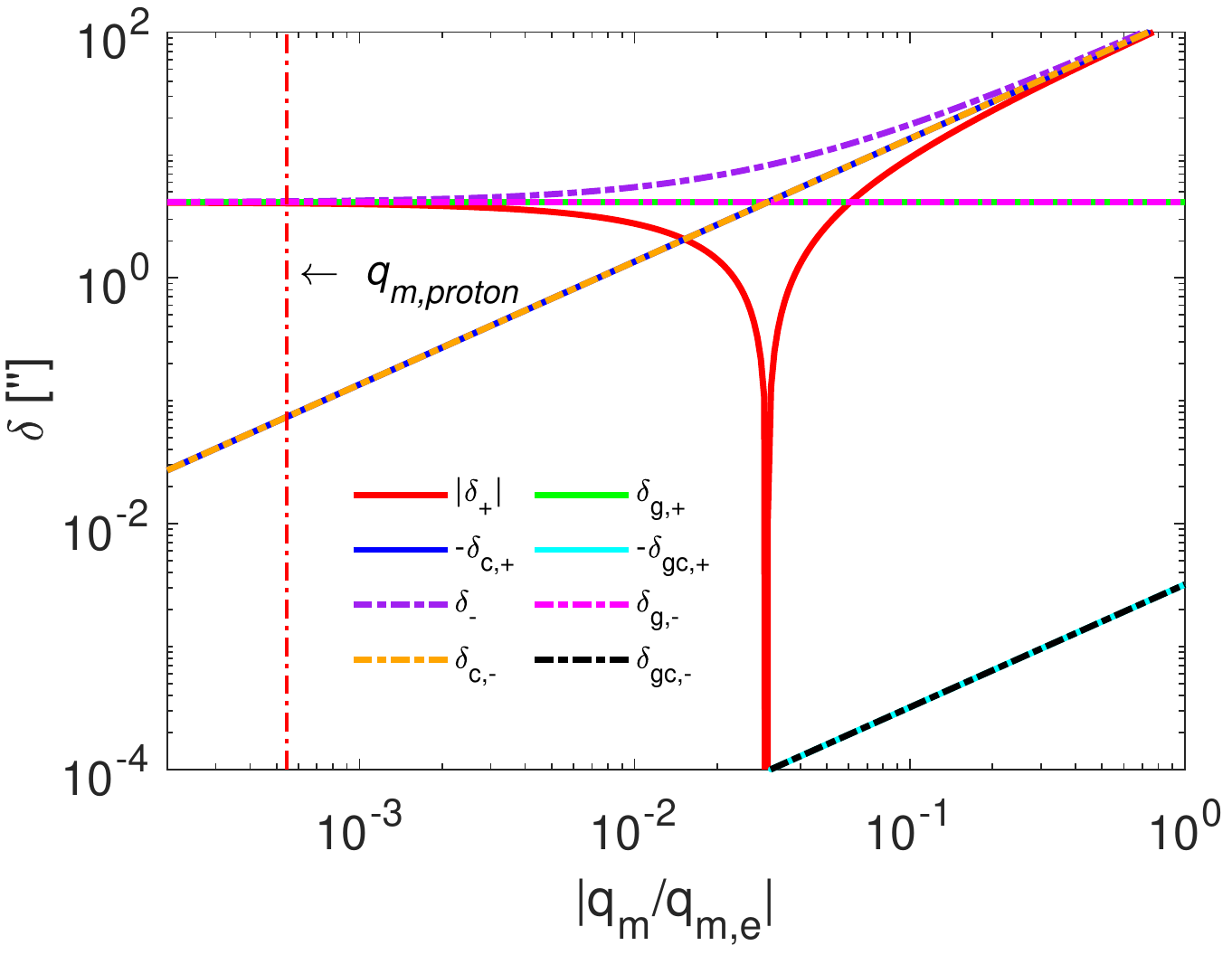}\\
(b)
\caption{Dependence of the total deflection $\delta$, the gravitational deflection $\delta_g$ and electrostatic deflection $\delta_c$ on (a) $Q$ with $q_m$ of proton and electron and on (b) $q_m$ with $Q=\sqrt{Q_{eq}Q_{mag}}$. In all plots, $\alpha=M^2/2,~r_S=r_O=8.1$ [kpc] and $b=1^{\prime\prime}r_O$ are used.}
    \label{fig:deltainqqmalpha}
\end{figure}

\begin{figure}[htp]
\includegraphics[width=0.4\textwidth]{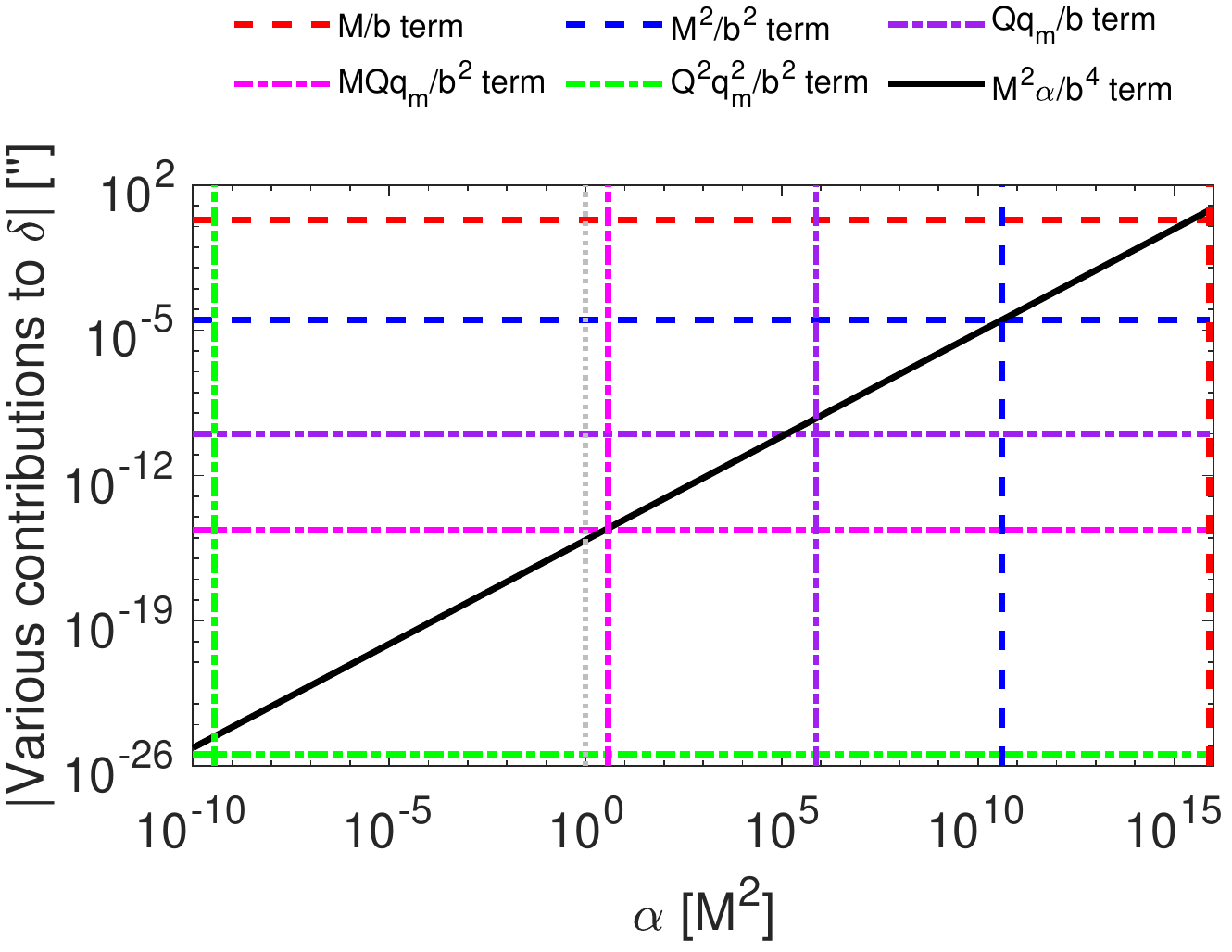}    \caption{Absolute value of various terms of Eq. \eqref{Sfangle} as $\alpha$ varies (see legend for labeling of the terms). All terms except the one containing $\delta_7$ are independent of $\alpha$ and therefore horizontal in this plot. See text for the vertical lines. $Q=Q_{eq}$ is used and other parameters are the same as in Fig. \ref{fig:deltainqqmalpha}.}
\label{fig:deltainqqmalpha_al}
\end{figure}

\begin{figure}[htp]
\includegraphics[width=0.4\textwidth]{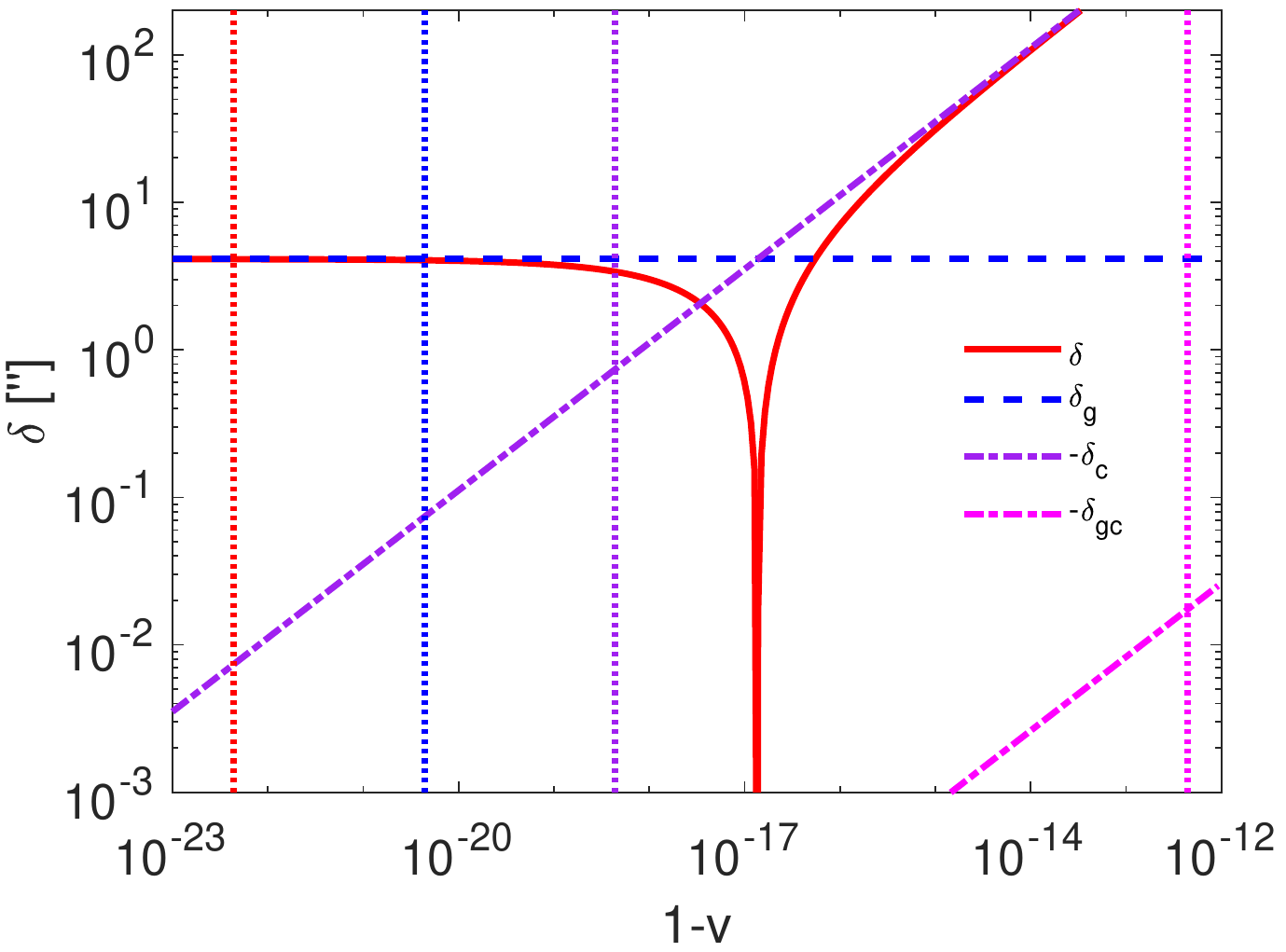}    \caption{Dependence of $\delta_g,~\delta_c$ and $\delta$ on $1-v$. The vertical lines from left to right correspond to the proton energy of $10^{20},~10^{19},~10^{18},~10^{15}$ [eV] respectively. $Q=100Q_{mag}$ is used and other parameters are the same as in Fig. \ref{fig:deltainqqmalpha}.}
\label{fig:deltainqqmalpha_v}
\end{figure}

We will choose the Sgr A* SMBH as the lens and then study the deflection angle of ultra-high energy (UHE) protons and other charged particles in the cosmic rays. Using its mass of $4.1\times 10^6 M_\odot$ and distance of 8.1 [kpc] \cite{Abuter:2018drb}, $10^{19}$ [eV] for proton energy (which fixes its velocity $v$) and an impact parameter $b$ corresponding to an apparent angle of $\mathcal{O}(1^{\prime\prime})$, the dependence of the deflection angle on $Q, ~q_m,~\alpha$ and $v$ are plotted in Fig. \ref{fig:deltainqqmalpha} to Fig. \ref{fig:deltainqqmalpha_v}.

In Fig. \ref{fig:deltainqqmalpha} (a), the effect of the charge $Q$ on the deflection \eqref{Sfangle}, the gravitational deflection $\delta_g$, the electric deflection $\delta_c$, and the coupling $\delta_{gc}$ are all plotted for UHE protons and electrons. It is seen that for the entire range of $Q$ considered, i.e. $[0,Q_l]$, the gravitational deflection of a proton, $\delta_{g,p}$, is almost (although not exactly) constant as $Q$ increases,  because $Q_l/Q_{ext}\ll 1$. In contrast, the magnitude of the electrostatic deflection $\delta_{c,p}$ increases almost linearly as dictated mainly by the first term of Eq. \eqref{eq:deltacoul}. The gravitational-electric coupling term $\delta_{gc}$ although also increases as $Q$ increases, its magnitude is smaller than that of $\delta_c$ by a factor of $M/b$, and therefore negligible in the competition of $\delta_g$ and $\delta_c$. For protons, the electrostatic repulsion is smaller than the gravitational attraction until $Q/Q_{mag}\approx 5.58\times 10^3$ or $Q=1.35\times 10^{19}$ [C], beyond which the total deflection becomes negative and the proton is pushed away from the central lens. Clearly, although this charge is much larger than the magnetically induced charge of the Sgr A* SMBH $Q_{mag}\approx 2.42\times 10^{15} $ [C], it is still much smaller than its extreme limit $M$. Therefore observing the deflection or GL of such protons by the Sgr A* BH might offer a chance to constrain its charge. 
On the other hand, if it was the electron signal that experience such deflections, then the electrostatic and gravitational interaction will cause the same amount of deflection when $Q$ is even smaller, at about $3.04Q_{mag}$. 

Fig. \ref{fig:deltainqqmalpha} (b) shows the deflection angle for charges with different $q_m$. Although the most probable signal candidates are UHE protons, for completeness we still allow the $|q_m|$ to range from roughly $ \frac{|e|}{2m_p}$ for heavy ion to $\frac{|e|}{m_p}$ for protons/antiprotons and then further to $\frac{|e|}{m_e}$ for electron/positron \cite{Lipari:2016vqk}. Note that to be conservative, we set the charge of the BH to a relatively large value $Q=100Q_{mag}$ in this plot. It is seen that unlike Fig. \ref{fig:deltainqqmalpha} (a), the gravitational deflection $\delta_g$ here is exactly a constant with respect to the change of $q_m$. Again, the magnitude of the electrostatic deflection $\delta_c$ grows linearly as $q_m$ increases. For our primary signals, UHE protons and possibly heavier ions (He, C, O etc.) \cite{Lipari:2016vqk}, it is seen that because their electrostatic interaction with the lens is repulsive, the deflection angles $\delta_c$ are quantitatively negative, as expected. More importantly, even for this relatively large $Q$, the size of $\delta_c$ for these signals are always smaller than their gravitational deflection $\delta_g$ by a factor of about 1/55.8 for (anti)protons and roughly 1/112 for heavier ions. For electrons/positrons in cosmic rays, it is generally expected that the innergalactic magnetic field will bend their trajectories more strongly than any of the gravitational deflection $\delta_{g,e}$ or electrostatic deflection $\delta_{c,e}$ considered here. However, if we were to compare $\delta_{g,e}$ and $\delta_{c,e}$, then as seen from the plot, this BH with a charge of $100Q_{mag}$ can even result in a larger $\delta_{c,e}$ than $\delta_{g,e}$, roughly by a factor of 32.9. Again, for all considered signals, their gravitational-electrostatic coupled deflections $\delta_{gc}$ are much smaller than the corresponding electrostatic deflection. 

For the coupling constant $\alpha$ of the GB term , currently, there is no theoretical constraint on its value except its dimension is of $[M^2]$. Its effect on the deflection angle is only through the fourth order term in $1/b^4$, i.e., the term containing $\delta_7$ in Eq. \eqref{Sfangle}. It is clear that this term is linear to $\alpha$, with a positive $\alpha$ decreasing the deflection angle while a negative one increasing it. In Fig. \ref{fig:deltainqqmalpha_al}, this term is plotted as a function of $\alpha$ for $Q=Q_{eq}$ and then compared with other terms of Eq. \eqref{Sfangle}.  
Comparing to the leading order term containing $\delta_1$, contribution of $\delta_7$ term to the deflection will be weaker if $|\alpha|<b^4/M^2$ (this value is marked by the red dashed vertical line in Fig. \ref{fig:deltainqqmalpha_al}). It will even be weaker than the second order term proportional to $M^2/b^2$ if $|\alpha|<b^2$ (blue dashed vertical line). However, since $Q$ itself is very small comparing to $Q_{ext}=M$, the contribution from $\alpha$ might be larger than that from terms involving $Q$ and $q_m$, depending on the exact relation between these parameters. Comparison of terms containing $\delta_7$ with those containing $\delta_2,~\delta_4$ and $\delta_5$, one can see that
when $\displaystyle \alpha\approx  Qq_m\sqrt{1-v^2}b^3/M^2,~Q^2q_m^2(1-v^2)b^2/M^2 ,~Qq_m\sqrt{1-v^2}b^2/M$ (marked by purple, magenta, green dot-dashed vertical lines) respectively, the contribution to the deflection angle from the parameter $\alpha$ would be comparable to the corresponding electrostatic contributions. The first of these three values is close to $bM$ and the second is close to $M^2$ (gray vertical line) while the last is only about $10^{-10}M^2$ for the given choice of other parameters. Therefore although the parameter $\alpha$ only starts to appear from the $1/b^4$ order, its contribution can be comparable to some of the electrostatic deflections, even if $\alpha$ is only at the order of $M^2$ or even much smaller. Lastly, in Fig. \ref{fig:deltainqqmalpha_al} the $\delta_3$ contribution to the deflection was not plotted because for the chosen parameters, this term is much smaller than all terms plotted in this figure, even smaller than the value of the $\alpha$ term at $\alpha=10^{-10}M^2$.

Finally for the effect of signal velocity, the most prominent feature recognizable from Eq. \eqref{Sfangle} is that both the electrostatic deflection $\delta_c$ and the gravitational-electrostatic deflection $\delta_{gc}$ are proportional to $\sqrt{1-v^2}$ or its positive powers and therefore vanish as $v\to 1$.  In other words, the higher the energy of the charged particle, the less electrostatic and gravitational-electrostatic deflections it will experience. In comparison, the gravitational deflection $\delta_g$ is nonzero even for null signals. This feature is illustrated in Fig. \ref{fig:deltainqqmalpha_v}, where we plot $\delta_c,~\delta_g$ and the total $\delta$ for different velocities/energies of the cosmic ray proton, assuming the charge of the BH is still given by the value $100Q_{mag}$. It is seen that for UHE protons with energy greater than $10^{19}$ [eV], $\delta_c$ is smaller than $\delta_g$ by at least a factor of $\sim 1/55.8$. Around $10^{17}$ [eV], the protons' electrostatic deflection becomes comparable to the gravitational one and they are of opposite directions. If the energy/velocity continue to decrease, then the deflection of the electrostatic force will bend the signal away from the lens. Also note that the gravitational-electrostatic coupling deflection $\delta_{gc}$ in this parameter setting is a factor $M/b$ smaller than $\delta_c$, but their way of dependence on $v$ are roughly the same. 

\section{conclusion} \label{conclusion}

In this paper, we have studied the deflection of a CMP by a charged BH in a novel four-dimensional EGB gravity using the GB theorem. In order to use this theorem, we constructed the Jacobi space $(\mathcal{M},J_{ij})$ as the background space, in which the motion of a CMP follows the geodesic. This equation is solved in the weak field approximation via iterative method. 

After integrating the Gaussian curvature, the result of the deflection angle of a CMP in the EGB BH spacetime is given in Eq.~\eqref{Sfangle}, which also takes into account the finite distance effect.  From this result, various limits (null limit, neutral limit, infinite distance observer/source limit, and RN limit) were obtained and compared with known literature. Moreover, we find that a positive coupling constant $\alpha$ decreases the deflection angle, while a negative $\alpha$ increases it. 

Using Eq. \eqref{Sfangle}, we modeled the Sgr A* SMBH as the EGB BH and studied its deflection to charged cosmic ray, primarily to UHE protons. Its found that the electrostatic deflection of proton with energy $10^19$ [eV] is smaller (or larger) than the gravitational deflection if the BH charge  is smaller (or larger) than $5.58\times 10^3Q_{mag}$. This might offer a new way to constrain the Sgr A* SMBH charge if the angular resolution of UHE protons can be significantly improved in the future. 

It is worthwhile to mention that the charged black hole solution in four-dimensional EGB gravity has the same form as the solutions in the gravity with a conformal anomaly~\cite{cai2010}, and the gravity with quantum corrections~\cite{Tomozawa,Cognola}. Therefore, after proper re-interpretation of the parameters, the deflection angle obtained in this paper is also the deflection angles in these theories. Therefore neither the method nor the results obtained in this work has to rely on the validity/importance of the 4D EGB metric.  Essentially, this is a methodological work that provides an elegant paradigm for studying the weak gravitational deflection of charged particles in charged
SSS spacetimes. It is also straightforward to apply the method developed here to other charged SSS spacetimes. 

\acknowledgements

The authors thank Haotian Liu for the illustration of Fig. \ref{fig:lenselec} and Dr. Nan Yang for helpful discussions.

\end{document}